\documentclass[amsmat,amssymb,amsfonts,aps,prb,twocolumn,showpacs]{revtex4}

\usepackage{graphicx}
\usepackage{dcolumn}
\usepackage{bm}
\newcommand{\beq}{\begin{equation}}  
\newcommand{\eeq}{\end{equation}}  
\newcommand{\beqa}{\begin{eqnarray}}  
\newcommand{\eeqa}{\end{eqnarray}}

\begin{document}

\title{Spin dynamics of current driven single magnetic adatoms and molecules}

\author{F. Delgado,  and J. Fern\'andez-Rossier}
\affiliation{Departamento de F\'{\i}sica Aplicada,
Universidad de Alicante, San Vicente del Raspeig, 03690 Spain}

\date{\today}

\begin{abstract}
A scanning tunneling microscope can probe the inelastic spin excitations of a single magnetic atom in a surface via spin-flip assisted tunneling in which transport electrons exchange spin and energy with the atomic spin.    If the inelastic transport time, defined as the average  time elapsed between two  inelastic spin flip events,  is shorter than the atom spin relaxation time,  the STM current can drive the spin out of equilibrium.  Here we model this process using rate equations and a model Hamiltonian that  describes successfully spin flip assisted tunneling experiments, including a single Mn atom, a Mn dimer and Fe Phthalocyanine molecules.  When the STM current is not spin polarized, the non-equilibrium spin dynamics of the magnetic atom  results in non-monotonic $dI/dV$ curves.  In the case of spin polarized STM current,  the spin orientation of the magnetic atom can be controlled parallel or anti-parallel to the magnetic moment of the tip. Thus, spin polarized STM tips can be used both to probe and to control the magnetic moment of a single atom.  
\end{abstract}

 \maketitle

\section{Introduction\label{intro}}

A single magnetic atom is arguably the smallest system where the spin can be used to  store classical and/or quantum  information.
Therefore,  there is great interest in probing and manipulating the spin state of a single atom or a single molecule in a solid state environment.   Examples of this are single Phosphorous donors   in Silicon,\cite{Kane_Nature_1998}
 nitrogen-vacancy centers in 
diamonds, \cite{Jelezko_Gaebel_prl_2004,Childress_Gurudev_science_2006,Hanson_Dobrovitski_science_2008,Neumann_Mizuochi_science_2008} single Mn atoms in II-VI \cite{Leger_Besombes_prl_2006,Besombes_Leger_prb_2008} and III-V \cite{Kudelski_Lemaitre_prl_2007} semiconductors, and single magnetic adatoms in surfaces\cite{Heinrich_Gupta_science_2004,Hirjibehedin_Lutz_Science_2006,Hirjibehedin_Lin_Science_2007,Otte_Ternes_natphys_2008,Chen_Fu_prl_2008,Krause_Bautista_Science_2007,Meier_Zhou_Science_2008,Wiesendanger_revmod_2009,Tsukahara_Noto_prl_2009,Fu_Zhang_prl_2009,Brune_Gambardella_sursci_2009,Zhou_Wiebe_natphys_2010}.  Whereas in most cases the spin of the single atom is probed by optical means,  the possibility of coupling the spin of as single atom to an electrical circuit is particularly appealing.

Tremendous recent experimental progress  has made it possible to probe the spin of a single and a few atoms deposited in conducting surfaces by means of scanning tunneling 
microscopes.\cite{Heinrich_Gupta_science_2004,Hirjibehedin_Lutz_Science_2006,Hirjibehedin_Lin_Science_2007,Otte_Ternes_natphys_2008,Chen_Fu_prl_2008,Krause_Bautista_Science_2007,Meier_Zhou_Science_2008,Wiesendanger_revmod_2009,Tsukahara_Noto_prl_2009,Fu_Zhang_prl_2009,Brune_Gambardella_sursci_2009,Zhou_Wiebe_natphys_2010}
There are two complementary techniques that afford this:  spin polarized STM and spin flip inelastic electron
 tunnel spectroscopy (IETS). 
The working principle of spin polarized STM is spin dependent 
magneto resistance,\cite{Slonczewski_PRB_1989}  similar to that of tunnel magneto resistance junction: tunneling between two spin polarized conductors depends on the relative orientation of their magnetic moments.  Control of the spin orientation of either the tip or the substrate affords spin contrast STM imaging.\cite{Wiesendanger_revmod_2009}

 In the case of   spin flip IETS, electrons tunnel from the  STM tip to the surface (or vice versa), and exchange their spin with the atom, so that they produce a spin transition,
whose energy is provided by  the bias voltage. Thus, a new conduction channel opens when the bias voltage is made larger than a given spin transition [see Fig. 1(c),1(d)]. This results in a step in the conductance as a function of bias and permits to determine the energy of the spin excitations, and how they evolve as a function  of an applied magnetic
 field.\cite{Heinrich_Gupta_science_2004,Hirjibehedin_Lutz_Science_2006,Hirjibehedin_Lin_Science_2007,Otte_Ternes_natphys_2008,Chen_Fu_prl_2008,Tsukahara_Noto_prl_2009,Fu_Zhang_prl_2009,Rossier_prl_2009}   When the atom is weakly coupled to its environment, 
the spin is quantized,\cite{Canali_McDonald_prl_2000,Strandberg_Canali_natmat_2007}
 the spin transitions have sharply defined energies which can be described with a single ion spin Hamiltonian whose parameters can be inferred from the
 experiments.\cite{Heinrich_Gupta_science_2004,Hirjibehedin_Lutz_Science_2006,Hirjibehedin_Lin_Science_2007,Otte_Ternes_natphys_2008,Chen_Fu_prl_2008} 
This is the case of Mn, Co and Fe atoms\cite{Heinrich_Gupta_science_2004,Hirjibehedin_Lutz_Science_2006,Hirjibehedin_Lin_Science_2007,Otte_Ternes_natphys_2008} as well as  Fe and Co  
Phthalocyanines,\cite{Chen_Fu_prl_2008,Tsukahara_Noto_prl_2009,Fu_Zhang_prl_2009} all of them deposited on a insulating monolayer on top  of a metal. Remarkably, spin flip IETS does not require a spin polarized tip to extract information about the spin dynamics.

Both IETS and spin polarized STM  are based upon the fact that the spin state of the atom affects the transport electrons, yielding a spin-dependent conductance. Therefore, we must expect that  the transport electrons do affect the spin of the atom. This is the main theme of this  paper.  In the case of spin flip IETS, there are two relevant time scales. On one side,  the  inelastic transport time or {\em charge time}  $T_q$, which is 
defined as the average  time elapsed between two  inelastic spin flip events . On the other side, the magnetic atom {\em spin relaxation time}, $T_1$. In the $T_q>> T_1$ regime, the transport electrons always interact with a atomic spin in equilibrium with the environment. As a result, the occupation of the spin states is bias independent and the conductance is expected to have flat plateaus in between the inelastic steps.  In the $T_q <<T_1$ regime this is no longer the case, the current drives the atomic spin out of equilibrium, so that the occupations of the spin states are bias dependent.  As we showed in a previous work,\cite{Delgado_Palacios_prl_2010} for the case of a single Mn atom,  this results in a  modified conductance line-shape, with non-monotonic behaviour in between steps.  In this work we give an extended account of these effects, and consider also the case of Mn dimers and FePc.

Non-equilibrium effects become particularly appealing when either the tip or the substrate are also spin polarized. In this case we have current flow between two magnetic objects, which is expected to result in spin transfer torque.\cite{Slonczewski_JMMM_1996} It has been proposed theoretically,\cite{Delgado_Palacios_prl_2010} and independently  verified
 experimentally,\cite{Loth_Bergmann_natphys_2010}
that  the spin orientation of a single Mn atom can be controlled with a a spin polarized tip.
In this paper we provide a thorough analysis of this effect and extend our study to include the effect of an external magnetic field, as in the case of the experiments.\cite{Loth_Bergmann_natphys_2010}

\begin{figure}
\begin{center}
\includegraphics[angle=270,width=0.98\linewidth]{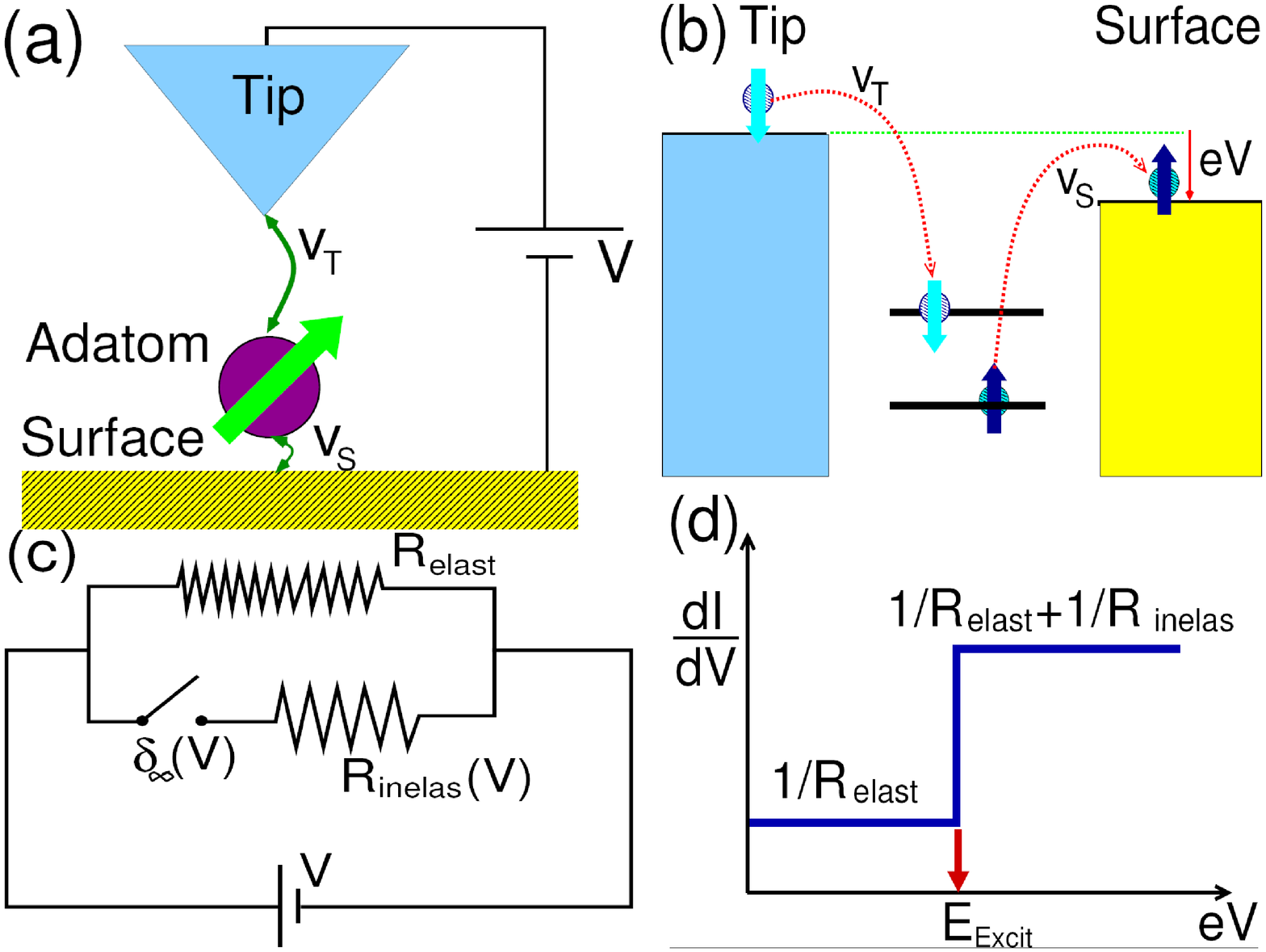}
\end{center}
\caption{ \label{fig1}(Color online). (a) 
Scheme of the proposed setup: a magnetic 
STM tip and a magnetic adatom on a insulating monolayer deposited on a metal. A current flows through the adatom when a bias voltage
is applied between tip and surface. (b) Energy level diagram of tip, atom and surface and typical microscopic process that gives rise to spin-flip tunneling. (c) Equivalent circuit that accounts for the increase of the conductance schematically shown in (d).}
\end{figure}

The results presented in this paper are based on a phenomenological spin-depedent tunneling 
Hamiltonian~\cite{Appelbaum_pr_1967,Rossier_prl_2009,Fransson_nanolett_2009,Delgado_Palacios_prl_2010,Fransson_Eriksson_prb_2010,
Zitko_Pruschke_prep} and are, in most instances, in agreement with existing experiments.  In the case of a single magnetic atom with spin $1/2$ it is possible to derive our  spin-dependent Hamiltonian from a single orbital Anderson model, by means of a Schrieffer-Wolff transformation.\cite{Anderson_prl_1966,Schrieffer_Wolff_pr_1966}  Within this picture, the spin-flip assisted tunneling events would correspond to inelastic cotunneling in the Anderson model.  In the single orbital Anderson model the spin-flip channel can dominate the elastic channel, which can be even zero in the so called symmetric case.   Further work\cite{Delgado_Rossier_prep} is in progress to generalize this picture  to the higher spin  case relevant to this paper.

The spin dynamics of current driven nanomagnets in the Coulomb Blockade  regime   has been thoroughly studied from the theory standpoint \cite{Efros_Rashba_prl_2001,Inoue_Brataas_prb_2004,Braun_Konig_prb_2004,Waintal_Parcollet_prl_2005,Romeike_Wegewijs_prl_2006,Lehmann_Loss_prl_2007,Elste_Timm_prb_2007,Rossier_Aguado_prl_2007,Fransson_prb_2008, Nunez_Duine_prb_2008,Delgado_Palacios_prl_2010}. The systems studied include magnetic grains\cite{Waintal_Parcollet_prl_2005,Michalak_Canali_prl_2006}, semiconductor\cite{Braun_Konig_prb_2004} and Mn doped quantum dots\cite{Efros_Rashba_prl_2001,Rossier_Aguado_prl_2007}, molecular magnets and magnetic molecules \cite{Romeike_Wegewijs_prl_2006,Misiorny_Barnas_prb_2007,Elste_Timm_prb_2007,Lehmann_Loss_prl_2007}.  In contrast to most of the theory work published to date, here we can model the current driven dynamics of a quantum spin whose Hamiltonian parameters  are accurately known from experiments \cite{Hirjibehedin_Lin_Science_2007}, making it possible to compare successfully theory and experiment.

The rest of the manuscript is organized as follows.  In Sec. II we present the model Hamiltonian for the magnetic atom(s),  the transport electrons, and their coupling, which accounts both for  spin-assisted tunneling and  Korringa like atomic spin relaxation due to exchange coupling with the electrodes.  The transition rates and non-equilibrium dynamics leading to the current are
analyzed in Sec. \ref{ratesmeq}. In Sec. III we present the results of current driven spin dynamics under the influence of non-magnetic tip in three cases: the single Mn adatom, the Mn dimer and the FePC molecule. In Sec. IV we discuss the case of a spin polarized tip and analyze in detail the case of a single Mn adatom. In Sec. V we present our main conclusions and discuss open questions.

%
\section{Theory}
\subsection{Hamiltonian\label{Hamiltonian}}

In this section we present the phenomenological  Hamiltonian, its microscopic justification, the rate equation approach for the atom spin dynamics, including both spin relaxation and spin driving terms,  and the calculation of the current. 
The system of interest is shown in Fig. \ref{fig1}(a). 
We use a  model Hamiltonian which describes the system of interest split in  3 parts: tip, substrate and the magnetic atom(s)
\cite{Delgado_Palacios_prl_2010}
\begin{equation}
{\cal H}= {\cal H}_{\rm T} + {\cal H}_{\rm S} + {\cal H}_{\rm Spin}+ {\cal V}.
\label{htot}
\end{equation}
The first two  terms describe the tip and surface surface:
\begin{equation}
{\cal H}_{\rm T} + {\cal H}_{\rm S}=\sum_{k,\sigma,\eta} \epsilon_{\sigma\eta}(k)  
c^{\dagger}_{k\sigma\eta}c_{k\sigma\eta},
\end{equation}
where $c^{\dagger}_{k\sigma\eta}$ creates and electron in electrode $\eta=T,S$, with momentum $k$ and spin $\sigma$ defined along the spin quantization axis,  $\vec{n}$.  Unless stated otherwise, we take $\vec{n}$ parallel to the magnetization of the tip, which is a static vector in our theory. 
Since we consider a non-magnetic surface, we have 
$\epsilon_{\sigma,S}(k)=\epsilon_S(k) $.
  All the results of this paper are trivially  generalized to the case of a non-magnetic tip and a magnetic surface.

The spin of the magnetic adatom(s) is (are) described with a single ion Hamiltonian, exchange coupled to other magnetic adatoms and to the transport electrons.\cite{Hirjibehedin_Lutz_Science_2006,Hirjibehedin_Lin_Science_2007,Otte_Ternes_natphys_2008,Chen_Fu_prl_2008,Rossier_prl_2009,Delgado_Palacios_prl_2010}
\beqa
\label{hchain}
&&{\cal H}_{\rm Spin}= \sum_{i}\left[D S_z^{'2}(i) + E\left(S_x^{'2}(i)-S_y^{'2}(i)\right)\right]
\crcr
&&+\frac{1}{2}\sum_{i,j,a} J_{i,j} S_a'(i).S_a'(j)+g\mu_B\sum_i \vec{S}'(i).\vec{B}.
\eeqa
The first term describes the single ion magneto-crystalline anisotropy, the second describes the inter-atomic exchange  couplings and
the third  corresponds to the Zeeman splitting term under an applied magnetic field $\vec{B}$.
Here the prime denotes that the spin quantization axis is chosen with $z'$ along the easy axis of the system, not along the magnetic moment of the tip, $\vec{n}$.  
This makes necessary to rotate ${\cal H}_{\rm Spin}$ when $\vec{n}$ is not parallel to the easy axis.
The value of the local spin $S(i)$, the magnetic anisotropy coefficients $D$ and $E$, and the exchange
coupling between atoms in the chain $J_{i,j}$, 
change from atom to atom and also depend on the substrate.\cite{Hirjibehedin_Lutz_Science_2006,Hirjibehedin_Lin_Science_2007,Otte_Ternes_natphys_2008,Barral_Weth_physb_2007,Rossier_prl_2009}
In the following, we  denote the eigenvalues and eigenvectors of ${\cal H}_{\rm Spin}$ as $E_M$ and $|M\rangle$ respectively.

We model the coupling of the magnetic chain with the reservoirs  with the following 
Kondo-like Hamiltonian:\cite{Appelbaum_pr_1967,Rossier_prl_2009,Fransson_nanolett_2009,Delgado_Palacios_prl_2010,Zitko_Pruschke_prep}

\begin{equation}
{\cal V}=
\sum_{\alpha,\lambda,\lambda',\sigma,\sigma',i} 
T_{\lambda,\lambda',\alpha}(i)
\frac{\tau^{(\alpha)}_{\sigma\sigma'}}{2}  \hat{S}_{\alpha}(i)
c^{\dagger}_{\lambda,\sigma} c_{\lambda'\sigma'}, 
\label{HTUN}
\end{equation}
where $i$ labels the magnetic atoms in the surface, 
 $\lambda=(k,\eta)$ labels the single particle quantum numbers of the transport electrons (other than their spin $\sigma$), and the index $\alpha$ runs over 4 values, $a=x,y,z$,   and  $\alpha=0$.
We use $\tau^{(a)}$ and $\hat{S}_{a}$ for the   Pauli matrices
and  the spin operators in the $\vec{n}$ frame, while $\hat{S}_{0}=\tau^{(0)}$ is the identity matrix.
$T_{\lambda,\lambda',\alpha}$ for $\alpha=x,y,z$ is the exchange-tunneling interaction between the localized
spin and the transport electrons, and potential scattering for $\alpha=0$.
Attending to the nature of the initial and final electrode, 
Eq. (\ref{HTUN}) describes four types of exchange interaction, two of which contribute to the 
current, the other two are crucial to account for the atom spin relaxation.

\subsection{Justification of the Hamiltonian}
 The phenomenological  spin models of Eqs. (\ref{hchain}) and (\ref{HTUN}) capture most of the experimental results, as we show below.  These models imply that that the magnetic atom is in a well defined  charge state except for  classically forbidden fluctuations that enable tunneling from the tip to the surface.   
Fig. \ref{fig1}(b) shows a typical level alignment in which the spin model can be applied. The basic condition is therefore, 
that the chemical potential of the electrodes must be far enough from the chemical potential of the central-quantized region. In this way, charge addition and charge removal are classically forbidden.
Although it is outside the scope of this work, we claim that the  quantum charge fluctuations that give rise to spin dependent tunneling are due to inelastic cotunneling. 
 In the case of spin $1/2$, the equivalence between the spin model, originally proposed by Appelbaum\cite{Appelbaum_pr_1967}  and a single site Anderson model was rigorously shown by Anderson\cite{Anderson_prl_1966} generalizing the Schrieffer and Wolff transformation\cite{Schrieffer_Wolff_pr_1966} to the case of a single site coupled to two reservoirs.  
 Within this picture, the atomic spin is exchanged coupled to the transport electrons and the magnitude of the exchange is given by
 $T_{\lambda,\lambda',\alpha=x,y,z}\simeq \delta^{-1} V_{\lambda}V_{\lambda'}$,
where $V_{k\eta}$ is the hybridization between the Anderson site and the single particle state $k$ in the electrode $\eta$, and $\delta$ is the energy difference between the Anderson level and the electrode Fermi energy. This is the so called Kinetic exchange.   Importantly, both electrode conserving and electrode non-conserving processes are included and their strengths are not independent, since they both depend on  hopping matrix elements $V_{\lambda}$ between the localized orbital in the atom and the extended orbitals in either the tip or the sample. Interestingly,  the Schrieffer and Wolff transformation\cite{Schrieffer_Wolff_pr_1966} also yields a spin-independent tunneling term which would yield the $\alpha=0$ contribution in Eq. (\ref{HTUN}), and it corresponds to the elastic tunneling contribution. Within this Anderson-Kondo picture, the strength of the $(\alpha=0)$ elastic channel  and that of the spin dependent channel $(\alpha=x,y,z)$ are comparable and, in the so called symmetric case, the elastic term vanishes identically. Thus, for spin $1/2$  case, this picture can account for the large strength of the inelastic signal. 
The generalization to higher spin case, relevant for the 
experiments,\cite{Heinrich_Gupta_science_2004,Hirjibehedin_Lutz_Science_2006,Hirjibehedin_Lin_Science_2007,Otte_Ternes_natphys_2008,Chen_Fu_prl_2008,Loth_Bergmann_natphys_2010} will be published elsewhere.\cite{Delgado_Rossier_prep}

Keeping these considerations in mind, and 
following Anderson,\cite{Anderson_prl_1966}  we assume that Hamiltonian (\ref{HTUN}) arises from kinetic exchange.
The momentum dependence of $T_{\lambda,\lambda',\alpha}(i)$
can have important consequences in  the conductance
 profile\cite{Merino_Gunnarsson_prb_2004} in an energy scale of $eV$, but it can be safely neglected in IETS. 
We thus parametrize 
\begin{equation}
T_{\eta,\eta',\alpha}(i)= v_{\eta}(i)v_{\eta'}(i)  {\cal T}_{\alpha},
\label{vfactors}
\end{equation}
where  $v_{\rm S}(i)$ 
and $v_{\rm T}(i)$ are  dimensionless factors that scale as the surface-adatom and tip-adatom
hopping integrals. Because kinetic exchange is spin rotational invariant we have ${\cal T}_{x}={\cal T}_y={\cal T}_z\equiv{|\cal T|}$.
  Thus, Eq. (\ref{HTUN}) implies that the spin-assisted tunneling and the atomic spin relaxations are both due to 
kinetic exchange, and Eq. (\ref{vfactors})  implies  that their amplitudes depend on the tip-atom and surface-atom tunneling matrix elements.  

\subsection{Rates and Master equation\label{ratesmeq}}
Our primary goal is to study transport and spin dynamics.  This is done considering  ${\cal V}$ as a perturbation to the otherwise uncoupled magnetic atom and transport electrons. The quantum spin dynamics is described by means of a master equation for the diagonal elements of the density matrix, $P_M$, described in the basis of eigenstates $|M\rangle$ of ${\cal H}_{\rm Spin}$.  The master equation is derived using the standard system plus reservoir technique,\cite{Cohen_Grynberg_book_1998} where the transport electrons act as a reservoir for the atomic spin(s).
The master equation\cite{Cohen_Grynberg_book_1998} reads
\begin{equation}
\frac{dP_M}{dt}=\sum_{M}P_{M'}W_{M',M}-P_M\sum_{M'}
W_{M,M'}\;,
\label{master}
\end{equation}
where $W_{M,M'}$ are the transition rates between the atomic spin state $M$ and $M'$. These rates can be written as  
$W_{M,M'}=\sum_{\eta,\eta'}W_{M,M'}^{\eta\to\eta'}$, where $W_{M,M'}^{\eta\to\eta'}$
are the scattering rates from an atomic spin state $M$ to $M'$ in which a 
quasiparticle electron goes from electrode $\eta$ to $\eta'$ as a result of exchange process.
They are given by:
\beqa
W_{M,M'}^{\eta \to \eta'}=\sum_{kk',\sigma\sigma'}\Gamma_{k\sigma M,k'\sigma 'M'}^{\eta\to \eta'}
f_{\eta}(\epsilon_{k\sigma})\left[1-f_{\eta'}(\epsilon_{k'\sigma'})\right]
\label{frates}
\eeqa
where $f_\eta(\epsilon)=1/(1+\exp{\left[-\beta(\epsilon-\mu_\eta)\right]})$ is
the occupation probability in electrode $\eta$ for electrons 
in equilibrium at chemical potential $\mu_\eta$ and temperature $T=1/( k_B \beta)$.
$\Gamma_{k\sigma M,k'\sigma 'M'}^{\eta \to \eta'}$ is the rate at which
an electron in lead $\eta$ with wavenumber $k$ and spin $\sigma$ is scattered into a lead 
$\eta'$ with wavenumber $k'$ and spin $\sigma'$, with the impurity spin undergoing a transition
between states $M$ and $M'$. Quantum rates $\Gamma$'s are calculated 
at the lowest order in the electrode-chain coupling using Fermi Golden rule
with the perturbation given by ${\cal  V}$ (see appendix A for  details):
\beqa
\Gamma_{k\sigma M,k'\sigma 'M'}^{\eta \to \eta'}&=&\frac{2\pi}{\hbar}\left|
 \sum_{\alpha,i} T_\alpha v_{\eta}(i)v_{\eta'}(i) S_{\alpha}^{M,M'}(i) 
 \right|^2
\crcr
&&\hspace{-0.9cm}\times
\delta\left(\epsilon_{\sigma\eta}(k)+E_M-\epsilon_{\sigma'\eta'}(k')-E_{M'}\right),
\label{qrates}
\eeqa
where we have defined the matrix elements
\begin{equation}
S_{\alpha}^{M,M'}(i) =\langle M|{S}_\alpha(i)|M'\rangle.
\label{transition-matrix}
\end{equation}
 The rates in Eq. (\ref{frates}) describe 3 types  of processes:
\begin{enumerate}
\item Elastic processes, $W_{M,M}^{\eta \to \eta'}$, in which the state of atomic spin remains unchanged and a transport electron is transfered from one electrode to another. These processes are responsible for the elastic current and have no effect on the spin dynamics. The rates of the elastic processes scale with $v_{T}^2v_{S}^2 T_0^2$. 

\item  Spin transitions $W_{M,M'}^{\eta \to \eta}$. In these, a spin transition in the atomic spin is produced due to the creation or annihilation of an electron hole pair either in the tip or in the surface.  These processes do not contribute to the current.  At very small temperature, the fastest process of this type is atomic spin relaxation: a spin transition from an excited state $E_M$ to a lower energy state $E_{M'}$ which results in the  excitation of an electron hole pair in one of the electrodes. This spin relaxation process is  very similar to the nuclear spin relaxation due to hyperfine coupling to  conduction electrons in metals and to Mn spin relaxation in diluted magnetic semiconductors due to itinerant carriers \cite{Besombes_Leger_prb_2008}.  At zero bias these processes dominate the atomic spin relaxation time $T_1$.  The rates of the spin transition processes scale like $v_{T}^4 {\cal T}^2$ and $v_{S}^4 {\cal T}^2$. 

\item Spin flip assisted tunneling $W_{M,M'}^{\eta \to \eta'}$. In these processes, which contribute both to inelastic current and to the dynamics of the atomic spin, a transport electron goes from electrode $\eta$ to $\eta'$ inducing a spin transition from state $M$ to $M'$.  The rates of the spin flip assisted tunneling processes scale like $v_T^2v_S^2 {\cal T}^2$.    

\end{enumerate}

The steady state solutions of Eq.(\ref{master})  depend, in general, on the Hamiltonian parameters, the temperature and the bias voltage.  We refer to the steady state solutions as $P_M(V)$. At zero bias, the steady state solutions are those of thermal equilibrium. At finite bias,  the $P_M(V)$ can depart significantly from equilibrium  depending on the relative efficiency of the transport assisted spin excitations and the spin relaxation.
Eq. (\ref{master}) does not include spin coherences. This approximation is good if the
 spin decoherence is faster than spin relaxation, which is known to be the case due to hyperfine 
coupling\cite{Gall_Besombes_prl_2009} in Mn atom. However, future work should address this point more carefully, in particular when the magnetization of the tip $\vec{n}$ is not parallel to the single ion easy axis.

\subsection{Relevant parameters}
The behaviour of the system is characterized by the  rates in Eq.(\ref{frates}), which depend on a number of physical quantities like the temperature, the bias voltage $V$, density of states at the Fermi Energy of tip and surface, $\rho_T$ and $\rho_S$, the tip-atom $v_T(i)$  and surface-atom $v_S(i)$ hoppings, the spin independent $T_0$ and spin dependent ${\cal T}$ couplings.  In this work we attempt to group the unknown parameters either in terms of dimensionless numbers or as experimentally accessible quantities. For that matter, we define the zero bias elastic conductance
\begin{equation}
g_0\equiv
\frac{\pi^2}{4}G_0 \rho_T \rho_S \left| T_0\right|^2\chi^2,
\label{g0}
\end{equation}
where  $G_0=2e^2/h$ is the quantum of conductance and
\begin{equation}
\chi=\sum_i v_T(i)v_S(i)
\end{equation}
is a parameter that quantifies the tip-surface transmission through the magnetic atoms. 
The density of states at the Fermi energy for spin $\sigma$ in the electrode $\eta$ are denoted by
 $\rho_{\eta\sigma}$ . 
We define the  spin assisted conductance $g_s$  as
\begin{equation}
g_S=\zeta^2 g_0
\end{equation}
where
\begin{equation}
\zeta=\frac{{\cal T}}{T_0}.
\label{z}
\end{equation}
is the ratio of the spin-flip assisted and elastic tunnel matrix elements.
We shall use spin polarization of the tip, defined as
\begin{equation}
{\cal P}_T= \frac{\rho_{T\uparrow}-\rho_{T\downarrow}}{\rho_{T\uparrow}+\rho_{T\downarrow}}.
\end{equation}
Another important parameter is  the ratio 
\begin{equation}
r(i)\equiv \frac{v_T(i)}{v_S(i)},
\end{equation}
which decreases as the tip is retracted from the surface.  In most instances, we shall have $r(i)<1$. 
 As a general rule,  the processes that drive the magnetic adatom out of equilibrium are proportional to $v_{\rm T}^2v_{\rm S}^2$ whereas the processes that cool the spin down (if $k_bT <|eV|$) are proportional to $v_{\rm T}^4+v_{\rm S}^4$. Thus, the non-equilibrium effects are higher as ${\cal R}\equiv r^2/(1+r^2)$ increases. 
%
 Therefore, current will be increased without changing the applied bias $V$. 
 In contrast,  the inelastic ratio $\zeta$ and the magnitude of the tip polarization ${\cal P}_T$ are not so easy to control. 

\subsection{Current\label{noneq}}
The calculation of  the rates for a tunnel event in which a transport electron goes from one electrode to the other, inducing a spin transition between states $M$ and $M'$ (where $M$ could be equal to $M'$ in the elastic channel), permits to obtain an expression for the current in terms of the steady state solutions of the master equation
\beqa
I_{S\to T}=e\sum_{MM'}P_M(V)\left( W_{M,M'}^{S\to T} -W_{M,M'}^{T\to S}\right),
\label{current0}
\eeqa
where 
$W_{M,M'}^{\eta\to\eta'}$ are the scattering rates from state $M$ to $M'$  induced by interaction with a quasiparticle which is initially in reservoir $\eta$ and ends up in $\eta'$, given in equation (\ref{frates}). 
We adopt the convention that positive bias voltage $V>0$ means {\em electrons flowing from tip to surface}, see Fig.~\ref{fig1}(b).
 Thus, we have:
\begin{equation}
eV=\mu_{\rm S}-\mu_{\rm T}
\end{equation} 
with $e$ the value of the electron charge with its sign.

\subsubsection{Current for non-magnetic tips}
The expression for the current in the case on non-magnetic tip and substrate 
can be written as the sum of two terms, elastic and inelastic, $I=I_0 + I_{IN}$ given by the expressions:\cite{Rossier_prl_2009}
 
\begin{eqnarray}
&&I_0=g_0 V,
\label{ielast}
\end{eqnarray}
where $g_0$ is given by Eq.(\ref{g0}) and
\begin{eqnarray}
I_{IN}=\frac{g_{S}}{G_0} \sum_{M,M',a}
 i_{-}(\Delta_{M,M'}+eV) \left| {\bf S}_{a,TS}^{M,M'}\right|^2 P_M(V)
\label{iinelast}
\end{eqnarray}
with
\begin{equation}
{\bf S}_{a,\eta\eta'}^{M,M'}\equiv \frac{1}{\chi}\sum_{i} v_{\eta}(i) v_{\eta'}(i) \langle M|{S}_a(i)|M'\rangle.
\label{transition-matrixW}
\end{equation}
Here we have introduced the current associated to a single channel with energy $\Delta_{M,M'}= E_M-E_{M'}$,  and bias $V$
\begin{equation}
i_{\pm}(\Delta+eV)= \frac{G_0}{e}\left({\cal G}(\Delta+eV) \pm {\cal G}(\Delta-eV)\right),
\end{equation}
with  ${\cal G}(\omega)\equiv \omega\left(1-e^{-\beta \omega}\right)^{-1} $. The curve $i_{-}$,  is odd in the bias, whereas $i_{+}$, relevant in the case of magnetic tips discussed below,  is even.  In contrast, $di_{-}/dV$ is even and $di_+/dV$ is odd. 
 In this non-magnetic tip case, the elastic current provides no information about the spin state whatsoever.  In contrast, the inelastic steps in conductance arise from Eq. (\ref{iinelast}) and permit to extract information about the spin transition energies, $\Delta_{M,M'}$ and spin matrix elements ${\bf S}_{a,TS}^{M,M'}$.
The basic effects of the elastic and inelastic terms in the conductance can be understood in terms of an equivalent electric circuit schematically shown in Fig.\ref{fig1}(c) and (d). 
For low enough voltage, the only channels that can conduct current are the elastic ones, while the inelastic channels remains close.  In this situation the (inelastic) switch is open. When the voltage 
is increased such as inelastic channels are open (switch is closed), these new channels contributes to the current, 
leading to a smaller resistance, see Fig.~\ref{fig1}(d).

\subsubsection{Current for magnetic tips}
In the case of a magnetic tip, the current has 3 contributions, $I=I_0+I_{MR}+ I_{IN}$.  This result is different from the non-magnetic case on two counts.  First, the elastic case has a magnetoresistive term, so that current is now proportional to the relative orientation of the average adatom spin and the magnetic moment in the tip $\vec{n}$:
\begin{eqnarray}
&&I_0+I_{MR}= g_0 \left[1+ 2\zeta \langle {\bf S}_{z,TS}\rangle {\cal P}_T\right]V,
\label{ielast-mag}
\end{eqnarray}
where
\begin{equation}
 \langle {\bf S}_{z,TS}\rangle=\sum_{M,i} P_M(V) v_{T}(i) v_{S}(i) \langle M|{S}_a(i)|M\rangle
\end{equation}
is the average magnetization along the $z$ axis, that we take parallel to the magnetic moment of the tip. As we show both below and in Ref. \onlinecite{Delgado_Palacios_prl_2010}, both $P_M(V)$ and the average atom magnetization depend on voltage. Importantly,  the magnetoresistive contribution to the elastic current  makes it possible to track changes in the single atom magnetization experimentally.

The second difference with the non-magnetic tip arises in the inelastic current, which is now given by the expression:\cite{Fransson_nanolett_2009,Delgado_Palacios_prl_2010}
\begin{eqnarray}
&&I_{IN}=\frac{g_{S}}{G_0} \sum_{M,M'}
\Big[ i_{-}(\Delta_{M,M'}+eV) \sum_a \left| {\bf S}_{a,TS}^{M,M'}\right|^2
 \nonumber \\
&&
+ {\cal P}_T i_{+}(\Delta_{M,M'}+eV){\bf \Xi}_{xy}(M,M')\Big]P_M(V).
\label{iinelast-mag}
\end{eqnarray}
The new term in the second line involves the matrix elements
\begin{equation}
{\bf \Xi}_{xy}(M,M')=2{\rm Im}\left[{\bf S}_{x,TS}^{M,M'}{\bf S}_{y,T,S}^{M',M} \right].
\label{defXi}
\end{equation}
As opposed to the standard inelastic current [Eq. (\ref{iinelast})], which gives rise to steps of equal height for positive and negative bias,   the ${\cal P}_T$ dependent term of the inelastic conductance, proportional to  $i_+$, yields  steps at the excitation energies of opposite sign as the polarity of the bias is reversed.  Both the elastic and inelastic term proportional to ${\cal P}_T$ can produce a $dI/dV$ which is not an even function of bias.

\section{Non magnetic  tip}
In this section analyze the implications of a non-equilibrium population distribution when a finite bias is
applied between tip and surface.   These effects will be more relevant when the current through the system increases (by increasing the coupling to the electrodes). Next, we will study these effects in three different systems:
  the Mn monomer, Sec.~\ref{monomer}, the Mn 
 dimer, Sec.~\ref{dimer} both  deposited on a Cu$_2$N surface and the iron Phthalocianine molecule, FePc, deposited on
an oxidized Cu$(1110)$ surface, Sec.~\ref{molecular}.


\subsection{Mn monomer\label{monomer}}
Let us consider first the case of a single Mn adatom in Cu$_2$N, which has been widely studied experimentally
\cite{Hirjibehedin_Lutz_Science_2006,Loth_Bergmann_natphys_2010} and 
theoretically. \cite{Rossier_prl_2009,Lorente_Gauyacq_prl_2009,Sothmann_Konig_2010,Rudenko_prb_2009,Lin_Jones_prep,Fransson_nanolett_2009,Fransson_prb_2008,Persson_prl_2009}
The spin of the Mn atom in this environment is $S=5/2$. 
The parameters of the single ion spin Hamiltonian have been determined experimentally to be  $D=-0.039$ meV,
$E=0.007$ meV,\cite{Hirjibehedin_Lin_Science_2007}  and  $g=1.98$.\cite{Hirjibehedin_Lutz_Science_2006}
  Since $E<<|D|$
we can limit our qualitative discussion to the case $E=0$, so that  the eigenstates of ${\cal H}_{\rm Spin}$
 are also eigenstates of $S_z$ (numerical simulations will be done with $E=0.007$ meV  and do not change qualitatively).
  In the absence of applied magnetic field and at temperatures much smaller than the zero field splitting $4|D|$, the equilibrium distribution is such that the two ground states, $S_z=\pm5/2$,  are equally likely and the average magnetization is zero. 
 At an energy of $4|D|$ above the ground state level, we find a couple of degenerate 
excited states, with  $S_z=\pm3/2$. Finally, the two states with $S_z=\pm1/2$ are found at $6|D|$.

\begin{figure}[t]
\includegraphics[height=0.55\linewidth,width=0.98\linewidth,angle=0]{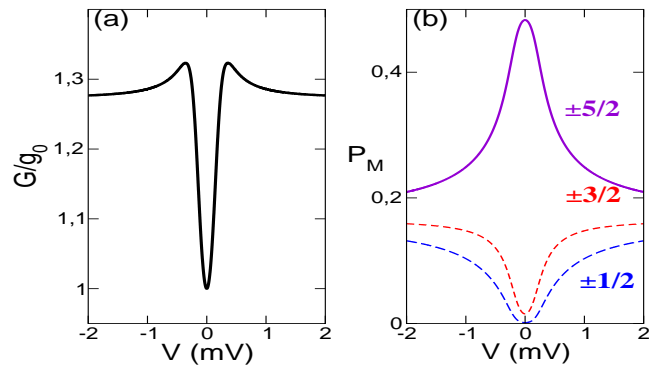} 
\caption{(Color online) (a) $dI/dV$, in units of $g_0$,  for single Mn in Cu$_2$N surface probed with a non-polarized tip 
for a fixed temperature $k_BT=0.5$K and $B=0$. {\bf (b)} Steady state populations of each eigenstate versus applied bias.
Here $\zeta=1$, $v_S=1$ and $v_T=1$.
}
\label{fig2}
\end{figure}

From the experiments, performed at low current,\cite{Hirjibehedin_Lutz_Science_2006} the  experimental 
$dI/dV$ lineshape is piecewise constant with two  steps symmetrically located   $eV=\pm 4D$. This is accounted for by the  equilibrium theory.\cite{Rossier_prl_2009}   As we show in Fig.~\ref{fig2}, and also in our previous work,\cite{Delgado_Palacios_prl_2010} non-equilibrium effects  modify  the $dI/dV$ lineshape. In particular,  the $dI/dV$ curve is not flat after the inelastic step
and it has a  small decay for $|eV|$ larger than the inelastic threshold. 
This  non-equilibrium effect    has been already observed 
experimentally.\cite{Otte_Ternes_natphys_2008,Loth_Bergmann_natphys_2010}.  Using the same theory with
a smaller tip-atom coupling (smaller $v_T$) results in $dI/dV$ lineshapes  identical to those  equilibrium calculations\cite{Rossier_prl_2009}.

 The non-monotonic $dI/dV$  can be explained as follows. As the bias goes across the inelastic threshold, $|eV|=4|D|$, there is a population transfer from the ground state doublet ($S_z=\pm 5/2$) to the first excited state doublet $S_z=\pm 3/2$.  
This can bee seen in Fig. \ref{fig2}(b). As soon as the population transfer to the first excited doublet takes place, a second inelastic channel opens:  the transition from the first to the second excited state doublet ($S_z=\pm 1/2$, whose energy is $2|D|$, smaller than the first step).  It turns out the intensity of the primary inelastic step ($\Delta=4|D|$),
 given by the matrix element $|\langle \pm 5/2| S^{\pm}|\mp 3/2\rangle|^2$, is larger than the intensity of the secondary transition ($\Delta=2|D|$).  Thus, the depletion of the primary transition in favor of the secondary one results in a decrease of the conductance.  In the case of FePc molecules, discussed  below, the secondary transition is stronger than the first one, resulting in an increase of the conductance after the first step.

\begin{figure}[t]
\includegraphics[height=0.8\linewidth,angle=-90]{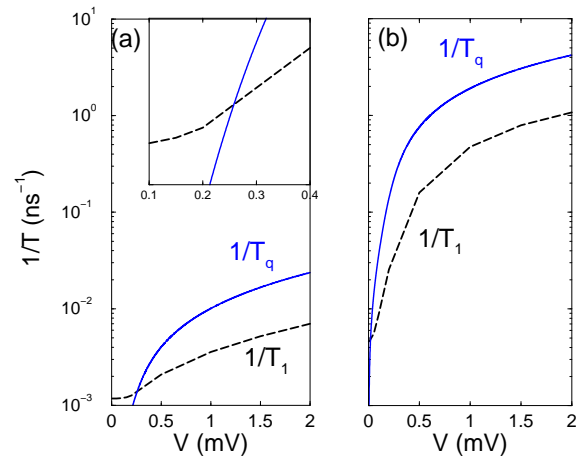} 
\caption{Inverse of the relaxation time of the spin state $S_z=+5/2$ (black dashed lines) and {\em charging time}
 $T_q^{-1}\equiv I_{IN}/e$ (blue solid lines),
versus the applied bias for two different couplings with the tip: (a) 
$v_T=0.1$, and (b) $v_T=1$. In both  cases, $T'=0.5K$, $v_S=1$ and $\zeta=1$.}
\label{timescale}
\end{figure}

   The non-equilibrium occupations can be understood as the balance between two  driving forces. Spin-flip assisted tunneling events heat the atomic spin,  delivering   energy of the order of $eV$  at a pace set by the inelastic current. 
  The steady state is reached when the heating power is exactly compensated by dissipation.  The later occurs via atomic spin relaxation due to exchange coupling to the tip and surface electrons. This process is enabled even at zero bias.   Interestingly, the steady state occupations can differ enormously from the zero bias thermal equilibrium. At $eV=2$meV,
 the  occupation of the ground state doublet is half of the one in equilibrium and barely twice the one of the higher energy spin levels, which are almost empty at zero bias.

  The inverse of the lifetimes of the two competing processes are shown in Fig. \ref{timescale}.
  There we show the relaxation rate $1/T_1$  of a magnetic spin state, i.e. $S_z=+5/2$, as a function of $V$. 
   When the tip is fully decoupled (no current through the system, ${\cal R}=0$), the spin relaxes in a time scale $T_1$
which is independent of the applied bias. For a small coupling, Fig.~\ref{timescale}(a), the relaxation rate increases by several orders of magnitude when the bias is increased. This effect is even more dramatic when the ratio $r$ approaches 1 [${\cal R}=1/2$, see Fig.~\ref{timescale}(b)]. 
In the weak coupling curve we can easily see the crossover from the equilibrium regime at low bias, where $T_1<<T_q$, to the non-equilibrium regime, for which $T_1>>T_q$.  In the other case the crossover occurs at a much lower voltage. 
To plot these curves we take a zero bias conductance $\sigma_0=0.75\mu$S for the $v_T=1$ case. 

%
%
\begin{figure}[t]
\includegraphics[angle=-90,width=0.8\linewidth]{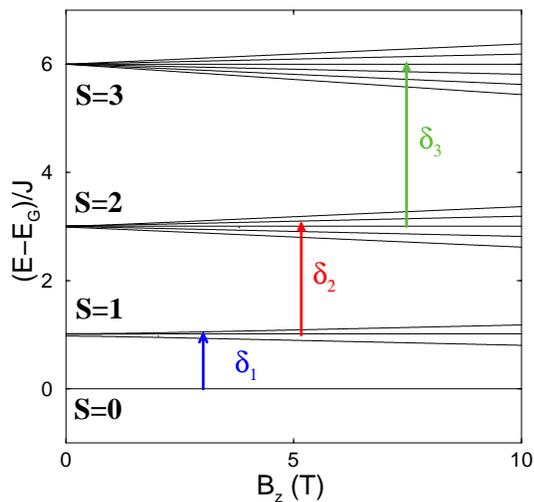} 
\caption{(a) Energy spectra corresponding to Hamiltonian (\ref{hchain}) for a Mn dimer
over a Cu$_2$N surface versus applied magnetic field.
 Spectrum is referred to the ground state energy and given in units of the exchange coupling  ($J=5.9$ meV). The magnetic
field is applied in the surface plane forming a $55º$ angle with the Cu-N direction.}
\label{fig4}
\end{figure}

\subsection{Dimer\label{dimer}}
\subsubsection{Non equilibrium effects}
From the discussion above,  the non-monotonic lineshape observed for the Mn monomer is related to non-equilibrium effects. Interestingly, correlation Kondo-like effects could also modify the lineshape\cite{Zitko_Pruschke_prep}. Given the fact that Kondo effect occurs in the case of a Cobalt atom deposited in the same surface,\cite{Otte_Ternes_natphys_2008} this type of effect can not be ruled out in the Mn monomer.  In contrast, the Mn dimer has a $S=0$ ground state\cite{Hirjibehedin_Lutz_Science_2006,Rossier_prl_2009,Rudenko_prb_2009} and
provides   an ideal system to test the non-equilibrium physics\cite{Loth_Bergmann_natphys_2010}.

The Mn dimer was studied experimentally under low current conditions by Hirjibehedin 
{\em et al.}\cite{Hirjibehedin_Lutz_Science_2006} and, more recently, under high current conditions by Loth
 {\em et al.}\cite{Loth_Bergmann_natphys_2010} They have observed a dramatic modification of the lineshape, 
which can be accounted for by our theory, as we show here. 
The Mn-Mn exchange  interaction in this system is antiferromagnetic. The fitting\cite{Hirjibehedin_Lutz_Science_2006} of the experimental results to the 
Hamiltonian model, Eq. (\ref{hchain}), gives a $J_{1,2}=J=5.9$, while $D,\; E$ and  $g$ are kept as for the monomer.

%
%
\begin{figure}
\includegraphics[height=1.\linewidth,width=0.78\linewidth,angle=-90]{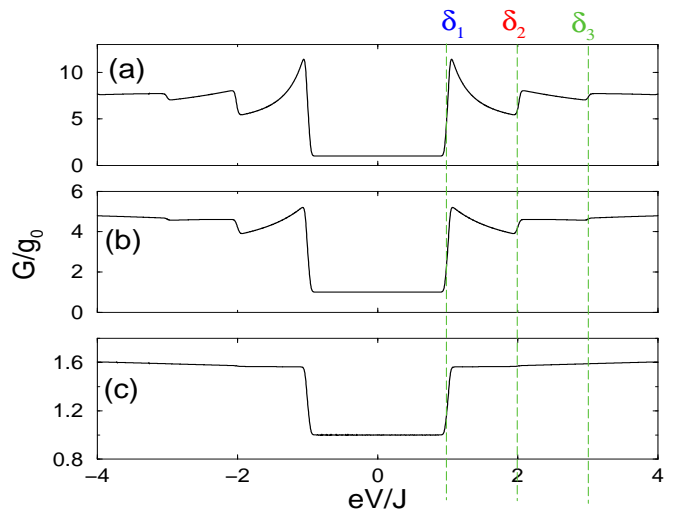} 
\caption{$dI/dV$ curve for the Mn dimer over a Cu$_2$N surface proved with a non-polarized tip. 
Each panel corresponds to a pair of $\left\{v_T(1),v_T(2)\right\}$ values: (a) $\{2,1.2\}$, (b) $\{0.4,0.1\}$ and (d) $\{0.01,0.001\}$.
 $T=0.6$K, $\zeta=1$, $v_S(i)=0.25$.}
\label{fig5}
\end{figure}

Fig.~\ref{fig4} shows the lowest energy spectra of the Mn dimer. Since $J>>|D|,E$,
the total spin $S$ is a good quantum number at zero order in $|D|/J$. Thus, the ground state is $S=0$, the first excited state $S=1$ and energy $J$, the second $S=2$ and energy $3J$ and the third $S=3$ and energy $6J$, all energies measured with respect to that of the ground state .  The $2S+1$ degeneracy of the $S>0$ multiplets is weakly lifted by the small anisotropy terms  $D$ and $E$. The allowed transitions induced by the exchange coupling (\ref{HTUN}), when the tip is more coupled to one of the two atoms,   satisfy  $\Delta S=\pm 1$. The lowest energy transitions are marked in Fig.~\ref{fig4} with vertical arrows at energies $\delta_i$,  with $i=1,2,3$.

The experimental results of the IETS show very different profiles as the current through the system is
 changed.\cite{Loth_Bergmann_natphys_2010} For low currents only the transition at energy $\delta_1 \approx J$ is
observed and  flat plateaus appear in the $dI/dV$ spectra before and after the inelastic step. This primary step corresponds to the transition from the $S=0$ ground state to the first excited state $S=1$.
As the current is increased, by reducing the tip-atom distance, additional steps appear at higher energies, corresponding to the transitions between the $S=1$ and $S=2$, and the $S=2$ and $S=3$ states, with energies  $\delta_2 \approx 2J$ and $\delta_3 \approx 3J$. In addition, the $dI/dV$ line shapes are not flat away from the steps either.
These results are  reproduced by our non-equilibrium theory. Fig.~\ref{fig5} shows the theoretical $dI/dV$ curve for 
three different couplings with the tip. When the tip is weakly coupled to the chain, Fig.~\ref{fig5}(a), the step corresponding to the $S=0\to S=1$ is clearly visible, while excitations from the $S=1\to S=2$ are quenched since the $S=1$ is only slightly populated.
When the coupling $v_T$ is increased (higher current), transitions $S=1\to S=2$ and $S=2\to S=3$ become possible for bias $|eV|>2J$ and $|eV|>3J$ respectively ( see
Fig.~\ref{fig5}(b)).  The new transitions are possible at high current due to a significant current induced occupation of the excited states $S=1$ and $S=2$  in the Mn dimer.  In contrast with the Mn monomer, the the excited state spin flip transitions energies $\delta_2$ and $\delta_3$  are larger than the primary spin transition, resulting in  in new steps in the spectra.  These experimental results, together with the theoretical interpretation,  provide strong evidence of the capability of the STM current to drive the spins of the magnetic adatoms.

\subsubsection{The  case of symmetric coupling}
\begin{figure}
\includegraphics[height=0.78\linewidth,width=0.578\linewidth,angle=-90]{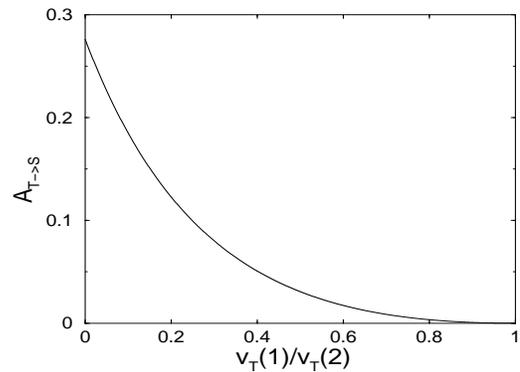} 
\caption{Transition amplitude  $A_{S\to T}$ versus the ratio $v_{{\rm T}}(1)/v_{{\rm T}}(2)$. 
 $T=0.6$K, $\zeta=1$, $v_S(i)=0.25$.}
\label{figstep}
\end{figure}

Whereas the results above are in very good agreement with the experimental data,\cite{Loth_Bergmann_natphys_2010}
 it is worth pointing out that this is only so if  we assume that the exchange assisted tunneling is stronger through one of the atoms.
However, it vanishes identically in the symmetric coupling case, $v_{{\rm T}}(1)=v_{{\rm T}}(2)$.
In Fig. \ref{figstep} we plot the height of the inelastic step, given by
$A_{S\to T}=\sum_{a}\sum_{M'} \left|{\bf S}_{a,TS}^{G,M'} \right|^2$,
 as a function of the lateral position of the tip across the dimer axis, as modeled by the ratio $v_{{\rm T}}(1)/v_{{\rm T}}(2)$. The inelastic step cancels identically when the tip is in the middle.  This prediction of the model is at odds with unpublished experimental data, which do not show a strong dependence of the inelastic current as the tip is moved along the Mn dimer axis.  From the theoretical point of view, the cancellation of the exchange assisted tunneling in the case of the Mn dimer symmetrically coupled to the tip arises from the fact that, in this particular case, the operator in the transition matrix element (\ref{transition-matrixW})  is the total spin of the dimer, and then the eigenstates of $S^2$ and $S_z$ are also eigenstates
of ${\cal V}$. As a result, the coupling Hamiltonian is diagonal, and no transitions are possible.  Notice that this problem is specific of the dimer. In the case of the monomer the observed spin transitions occur within states with the same $S=5/2$. In the case of  the trimer and longer chains the tip can not be coupled identically to all the atoms and the theory accounts for the data.\cite{Rossier_prl_2009}

There are several spin interactions other than the interatomic exchange that break the spin rotational invariance and could, in principle, solve the problem: the single ion anisotropy terms, $D$ and $E$,  the hyperfine coupling with the nuclear spin of the Mn, $I=5/2$, and the direct magnetic dipolar coupling. We have included them in our calculations, but they are much weaker than the dominant exchange, so that they do not change qualitatively the curve (\ref{figstep}).  Thus, even in spite of the apparent success of the perturbative approach using Eq. (\ref{HTUN}),  this particular result indicates the presence of additional terms in the Hamiltonian or the need to go beyond lowest order in perturbation theory . 
 Further work, going beyond the phenomenological theory is under way.\cite{Delgado_Rossier_prep}

\begin{figure}[t]
\vspace{0.5cm}
\includegraphics[angle=-90,width=1.\linewidth]{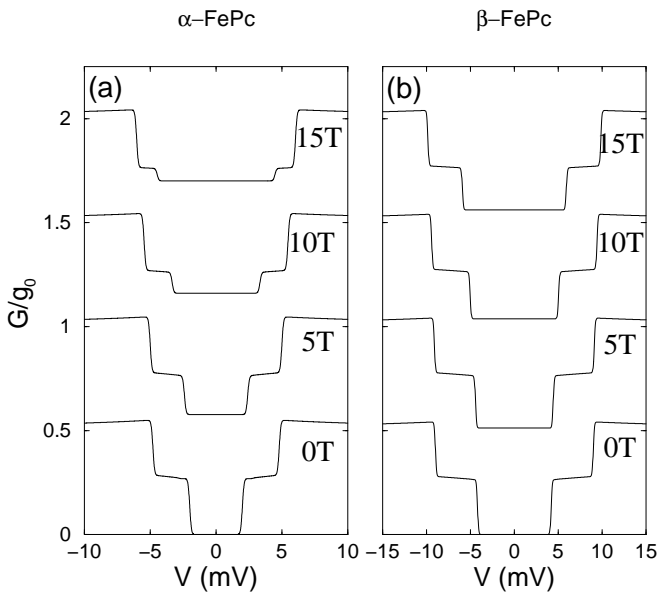} 
\caption{Differential conductance versus applied bias at zero magnetic field for the $ \alpha$ (a) and
$\beta$ (b) FePc on Cu$(110)(2\times1)-O$ as a function of applied bias for different magnetic field applied perpendicular to the
sample surface, as in Ref. \onlinecite{Tsukahara_Noto_prl_2009}.
Curves have been shifted by $0.5$ units for clarity. $T=0.5$K, ${\cal P}_T=0$, $v_S=1$, $v_T=0.2$ and $\zeta=1$.}
\label{fig7}
\end{figure}

\subsection{Magnetic molecules \label{molecular}}
As a final example of our non-equilibrium theory with non-magnetic electrodes, we consider 
the case of IETS through Iron  Phthalocyanine  (FePc) molecules,  deposited on 
 oxidized Cu  surface.\cite{Tsukahara_Noto_prl_2009}
   FePc are flat organic molecules with $C_4$ symmetry with a core made of a single Fe$^{2+}$ ion surrounded by 4 Nitrogen atoms embedded in Benzene groups. In gas phase, the crystal field of the ligands is high enough as to reduce the spin of $Fe^{2+}$ from $S=2$ (high spin) to $S=1$ (intermediate spin). Because of the $C_4$ symmetry of the gas phase, the single spin Hamiltonian of the molecule has $E=0$. 
 
According to the IETS data,\cite{Tsukahara_Noto_prl_2009} the symmetry is reduced when deposited on the oxidized surface. In particular,   
two adsorbed states ($\alpha$ and $\beta$) were experimentally observed with different spin excitations.\cite{Tsukahara_Noto_prl_2009} 
In both cases the spin excitations of the FePc could be assigned to  $S=1$ but the  anisotropy parameters,  determined from the
experimental differential conductance curves, varied in the two cases. For the $\alpha$ ($\beta$) configuration,
$D=-3.8\pm0.04$ meV ($D=-6.9\pm0.04$ meV), $E=1.0\pm0.01$ meV ($E=2.1\pm0.04$ meV) and  $g=2.3\pm0.02$ ($g=2.4\pm0.05$). The origin of this drastic change in anisotropy deserves further theory work. 

The $S=1$ single spin model can be solved analytically (see for instance appendix in Ref. \onlinecite{vanbrie_koenraad_prb_2008}).  With $D<0$, and  $E=0$, the ground state would be the $S_z=\pm 1$ doublet with energy $-D$ below the $S_z=0$ excited state. At finite $E$ the ground state doublet splits,   in bonding and anti-bonding combination of the states $S_z=\pm 1$. The splitting is $2E$. Thus, there are two spin transitions. The low energy one, with $\Delta=2E$ and $\Delta S_z=0$,  and the high energy one, with energy $|D|+E$ and $\Delta S_z= \pm 1$.   The magnetic field along the $z$ axis competes with the $E$ induced splitting of the ground state. As  $B$ increases, the $E$ induced mixing of the $S_z=\pm 1$ components decreases, and so it does the primary transition, which occurs  via $\Delta S_z=0$ events.

Fig. \ref{fig7} shows our non-equilibrium  theoretical $dI/dV$ results using the values of 
 $D$ and $E$ given above.  Our theory reproduces not only the evolution of the steps with the magnetic field,  but also the mild non-equilibrium features reported in Ref. \onlinecite{Tsukahara_Noto_prl_2009}. After the first (second) step the conductance has a small positive (negative) slope. In contrast,  the equilibrium theory,\cite{Gauyacq_Novaes_arXiv_2010} yields flat steps. The sign of the non-equilibrium slopes depends on the relative value of the inelastic channel strengths of the primary and secondary transitions.  At $eV=2E$, when the first excited state is populated, the secondary transition, with energy $|D|-E$ becomes possible, at finite temperature. Since this transition between excited states has a larger quantum yield, the  overall conductance increases.  The opposite scenario occurs in the second step.

\section{Spin polarized tip}
The results of the previous section give very strong support to the notion that tunneling electrons can drive the spin of the magnetic adatoms far from equilibrium.  Since we have been considering spin unpolarized tunneling electrons, these non-equilibrium effects can not result in a net spin transfer. 
From the theory standpoint, this should change dramatically in the case of spin-polarized transport electrons. 
As it was shown in a seminal work by Slonczewski,\cite{Slonczewski_JMMM_1996}
the back action of transport electrons on a magnetic moment
can be used to rotate the magnetization direction. This effect, known as spin-transfer torque, 
have been observed in nano-pillars of tens of nanometers\cite{Myers_Ralph_Science_1999} down to tiny  nanomagnets made of
100 atoms,\cite{Krause_Bautista_Science_2007} but still in the semiclassical domain. 
In a previous paper \cite{Delgado_Palacios_prl_2010} we modeled the spin dynamics of a single Mn atom under the influence of spin polarized current.  We found that the  if the tip was spin polarized, 
the spin polarized current would  result in a net spin magnetization of the magnetic adatom whose orientation relative to the tip moment would depend on the polarity of the bias, quite in agreement with the macroscopic spin transfer torque. 
In parallel to our work,  S. Loth {\em et. al.}\cite{Loth_Bergmann_natphys_2010} demonstrated experimentally the single atom spin transfer.

 In our work in Ref. \onlinecite{Delgado_Palacios_prl_2010} the origin of the tip magnetization was ferromagnetic order. In the experiment of Loth {\em et al.}, the spin polarized current is achieved by sticking a single Mn atom into the tip and applying a magnetic field to freeze its  spin fluctuations. The external magnetic field affects also the surface atom. The very different role played by the Mn in the tip and the Mn in the surface underlines the important role played by spin isolation. Whereas it is still possible to model the spin of the  Mn in the surface  as a quantized spin weakly  coupled to the surface electrons, this picture seems to break down for the case of the Mn in the tip, due to a combination of  charge transfer, Kondo coupling and very reduced spin lifetime.  Thus, the Mn in the tip acts as a spin filter for the transport electrons. Whereas this picture works qualitatively, we believe this issue deserves further work.\footnote{We acknowledge A. S. N\'u\~nez for this remark}

\subsection{Current induced spin switching\label{pumping}}

The flow of  spin polarized current through a single magnetic atom is expected to result in a transfer of a net spin into the atom. In the case of a single or a few magnetic atoms, where time reversal symmetry is not spontaneously broken at zero magnetic field , the equilibrium occupation of states with opposite $S_z$ is the same, resulting in a null average magnetization.  Spin polarized current changes this situation via spin-flip inelastic tunnel.\cite{Delgado_Palacios_prl_2010} The mechanism is the following. The dominant inelastic transitions  in the case of the Mn monomer are:
\begin{itemize}
\item Spin increasing  (SI) transition, for which the Mn spin goes from $S_z=-\frac{5}{2}$ to $S_z=-\frac{3}{2}$ and the transport electron goes from the high energy electrode with spin$\uparrow$ to the low energy electrode with spin  $\downarrow$.

\item Spin decreasing  (SD) transition, for which the Mn spin goes from $S_z=+\frac{5}{2}$ to $S_z=+\frac{3}{2}$ and the transport electron goes from the high energy electrode with spin$\downarrow$ to the low energy electrode with spin  $\uparrow$.

\end{itemize} 
In the case of spin unpolarized current, these two processes are equally likely and result in the depletion of the two states of the ground state doublet shown in Fig. \ref{fig2}(b). In the case of a spin polarized tip, the two processes are no longer equally likely, resulting in a net spin transfer from the spin current to the atomic spin. Let us consider the case where there are more $\downarrow$ than $\uparrow$ electrons in the tip. This means negative tip spin average   (i.e., ${\cal P_T}<0$) and positive tip magnetization.  When electrons go from the tip to the surface ($V>0$) , the   $SD$ processes are dominant, as a result of which the positive  $S_z$ states are depleted and a negative $\langle S_z \rangle$ is expected.   Thus, we expect that at  positive  bias (electrons going from tip to surface) the current co-polarizes the spin of the atom.

We now consider electrons going from surface to tip. Since the density of states of spin $\downarrow$ electrons is higher,  the $SI$ process is now more likely than the $ST$ one.  As a result, the negative $S_z$ states should be depleted, resulting in a positive atomic spin. Thus, we expect that $V<0$  (electrons going from surface to tip) the current counter-polarizes the spin of the atom.

\begin{figure}[!]
\begin{center}
\includegraphics[height=0.93\linewidth,width=0.55\linewidth,angle=-90]{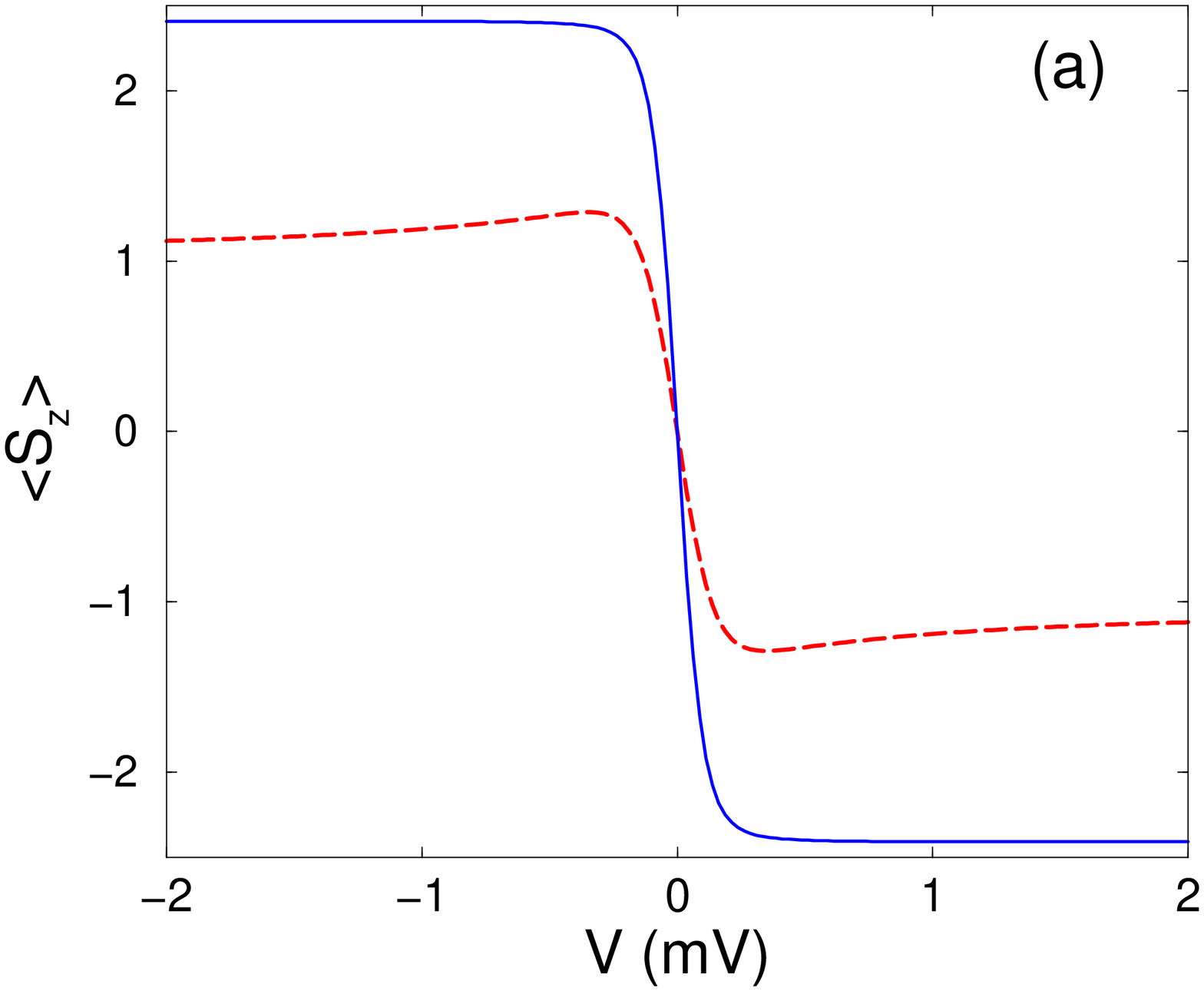}
\end{center}
\begin{center}
\includegraphics[height=0.93\linewidth,width=0.55\linewidth,angle=-90]{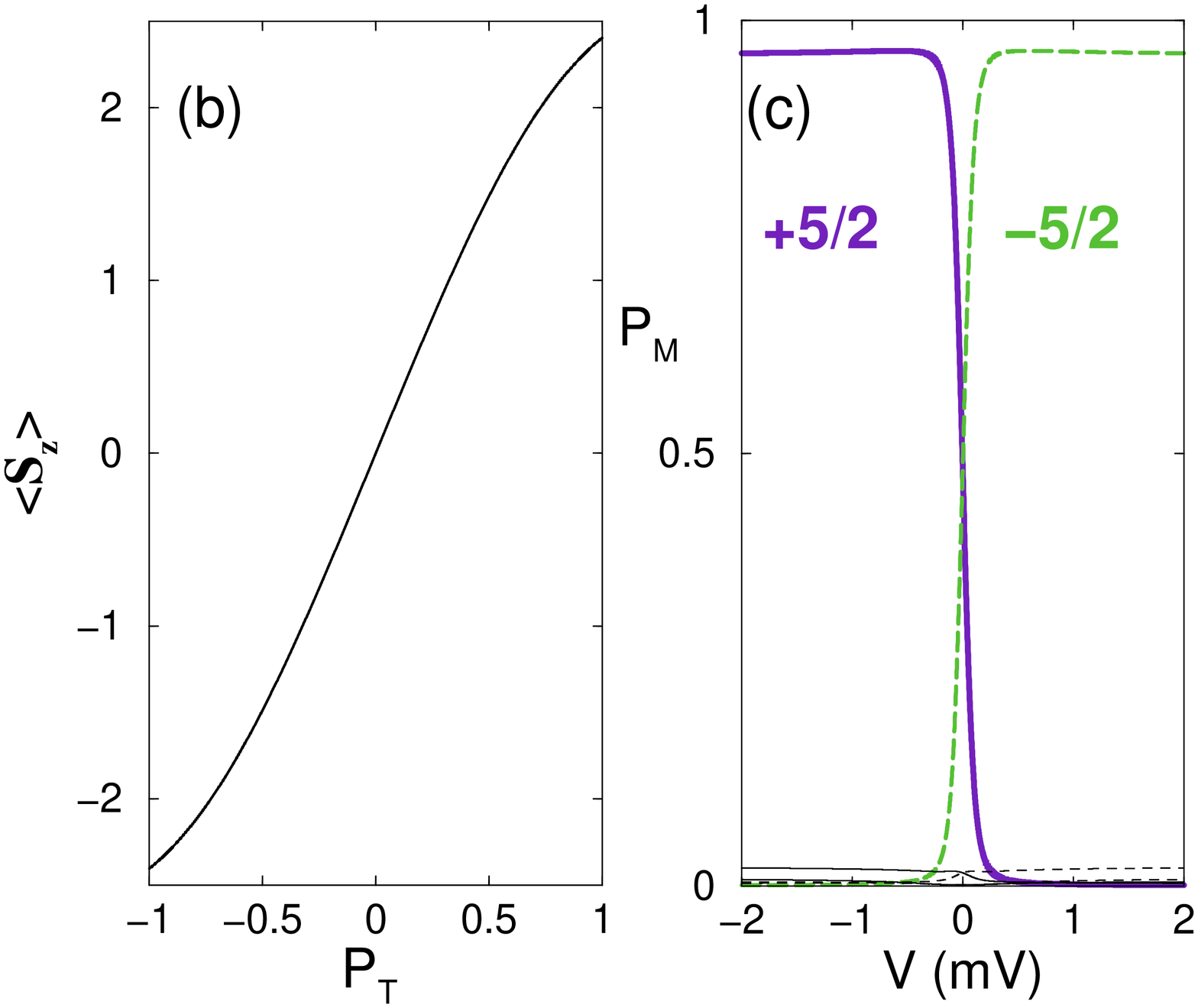}
\end{center}
\caption{(Color online)
(a) Average magnetization $<S_z>$ for the Mn monomer in Cu$_2$N surface probed with two differently  polarized tips,
 ${\cal P}_T=-1/3$ (red-dashed line) and ${\cal P}_T=-1$ (solid-blue line) versus applied bias.
(b) ``Universal curve'' of the magnetization versus
tip polarization (for positive applied bias).
(c) Steady state populations of each eigenstate versus applied bias. Solid lines for energy levels with $S_z<0$ and dashed lines 
for $S_z>0$.
Here $\zeta=0.5$, $v_S=1$, $T=0.5$K, $B=0$. and $v_T=0.7$.
}
\label{fig3pol}
\end{figure}

Our simulations confirm this scenario. We only consider the simplest case in which
the tip polarization is assumed parallel to the Mn easy axis -perpendicular to the Cu$_2$N surface. We consider first the case of zero  magnetic field.  We choose ${\cal P}_T<0$, because  is convenient for the discussion at finite positive field below. 
In Fig.~\ref{fig3pol}(a) we show the average atomic spin moment along the easy axis, as a function of the applied bias.  It vanishes at zero bias, reflecting the absence of spontaneous time reversal symmetry breaking of such a small system.  At finite bias the magnetic moment aligns with that of the tip when electrons flow from tip to surface ($V>0$),  and do exactly the opposite when the electrons flow from the surface to the tip ($V<0$).
Interestingly, the average atomic spin is finite even when $|eV|<4|D|$, the excitation energy. 
This is due to the existence of  thermally excited quasiparticles. However, the time necessary to drive the spin of the atom increases exponentially when  for $|eV|<4|D|$.\cite{Delgado_Palacios_prl_2010}

The average atomic spin  increases both with the applied voltage and the spin polarization of the tip, as given by ${\cal P}_T$.  The effect is null at ${\cal P}_T=0$, as it should, and it is maximal for half-metallic tips ${\cal P}_T=\pm 1$. 
For a fixed tip polarization the effect saturates at a certain voltage. Interestingly, the saturation magnetic moment depends only on the value of ${\cal P}_T$ and is quite independent of temperature and other parameters in the calculation. We discuss this universal behavior  in Sec. \ref{universal}.
The non-zero atomic spin polarization  reflects the bias induced breaking asymmetry of the  steady state  occupation of the two states of the ground state doublet $S_z=\pm 5/2$,  as shown in Fig.~\ref{fig3pol}(c). 
Notice the striking difference with the case of equilibrium, for which the  occupations of these two degenerate states are  identical. The steady state is reached thanks to the  competition between bias induced spin-transfer and  exchange induced spin relaxation discussed in the previous section.

\subsection{Effects of spin polarization on transport\label{polarization}}
\begin{figure}[!]
\includegraphics[angle=-90,width=0.89\linewidth]{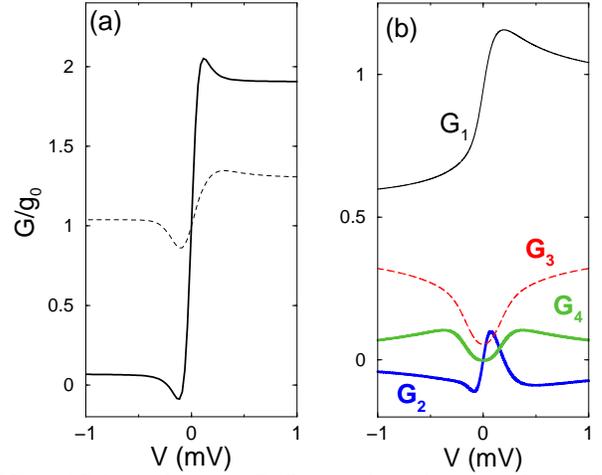} 
\caption{(Color online) (a) Differential conductance for the Mn monomer versus the applied bias for two different tip polarizations: ${\cal P}_T=-1/3$ (dashed line) and ${\cal P}_T=-1$ (solid line). (b) Each of the contributions to the $dI/dV$ for ${\cal P}_T=-1/3$: 
$G_1$(thin-black line), $G_2$ (thick-blue line), $G_3$ (thin-dashed line)
and $G_4$ (thick-green line) versus applied bias.  $T=0.5$K, $\zeta=0.5$ and $v_T=0.7$. Magnetization direction was fix parallel to the easy axis and no magnetic field was applied.}
\label{fig9}
\end{figure}

Importantly, the current induced polarization of the atomic spin can be detected through its influence on 
 the conductance of the system. The  simplest effect comes from the elastic   magnetoresistance: conductance is larger when spin polarization of tip and magnetic atom are parallel.   In the case discussed above, this results in a larger conductance at large positive bias than at large  negative bias.  
  At small bias, there are several competing effects. 
For simplicity let us consider the case of a single magnetic adatom.
We can write the differential conductance as 
\begin{equation}
G=G_1+G_2+G_3+G_4
\end{equation}
 where the different terms are obtained by deriving $I$ in Eqs. (\ref{ielast-mag}-\ref{iinelast-mag})
 with respect to bias and are shown in Fig.~\ref{fig9}(b):
 \beqa
G_1&=& g_0\left(1+2\zeta \langle S_z \rangle {\cal P}_T\right),
\crcr
G_2&=&2 g_0 V \zeta {\cal P}_T\frac{d \langle S_z\rangle}{dV},
\crcr
G_3&=& \frac{g_{S}}{G_0} \sum_{M,M'}P_M(V)
\Bigg[ \sum_a \left| S_a^{M,M'}\right|^2
\crcr
&&\hspace{-1.25cm}\times i_{-}'(\Delta_{M,M'}+eV) 
+ {\cal P}_T \Xi_{xy}(M,M')i_{+}'(\Delta_{M,M'}+eV)\Bigg],
\crcr
G_4&=&\frac{g_{S}}{G_0} \sum_{M,M'}
\Bigg[ i_{-}(\Delta_{M,M'}+eV) \sum_a \left| S_a^{M,M'}\right|^2
\crcr
&&\hspace{-1.25cm}
+ {\cal P}_T i_{+}(\Delta_{M,M'}+eV)\Xi_{xy}(M,M')\Bigg]\frac{dP_M(V)}{dV},
\label{conduct}
\eeqa
with $i_\pm'\equiv di_{\pm}/dV$. $\Xi_{xy}(M,M')$ is defined as in Eq. (\ref{defXi})
 but with without the
weighting factors. $G_1$ and $G_2$ ($G_3$ and $G_4$) correspond to the elastic (inelastic) contribution of the current. $G_1$  gives the dominant magnetoresistive contribution at large bias discussed above. $G_2$ gives a smaller contribution associated to the change of the average adatom spin as  a function of bias. This term is responsible of the non-monotonic decay of the conductance after the pronounced change induced by $G_1$. 
In the extreme case shown in Fig.~\ref{fig9}, corresponding to a half metallic tip,  $G_1$ is the dominant contribution. 
Finally, the two inelastic contributions, $G_3$ and $G_4$  peak close to the transition energies $\pm 4|D|$. 
$G_3$ corresponds to the inelastic conductance, as if the occupations $P_M(V)$ where bias independent, and $G_4$ is the contribution coming from the fact that $P_M$ do depend on the bias. In turn, both $G_3$ and $G_4$ have two contributions, one that is present for non-magnetic tip and another one proportional to the tip spin polarization ${\cal P}_T$.  

Experimentally it might be hard to disentangle $G_2$, $G_3$ and $G_4$, but not $G_1$ which provides a direct way to quantify the atomic spin at large bias, denoted by $\pm V_{\infty}$:
\begin{equation}
\frac{G_1(+V_{\infty})-G_1(-V_{\infty})}{G(V=0)}=  4\zeta  {\cal P}_T \langle S_z \rangle(+V_{\infty})
\label{G1vsSz}
\end{equation}
where $G(V=0)=g_0$ is the zero bias conductance and we have used the fact that, at zero magnetic field, 
$ \langle S_z \rangle(+V_{\infty})=  -	\langle S_z \rangle(-V_{\infty})$. Since $G_1$ can be the dominant contribution, replacing $G_1$ by the total $G$ in  equation (\ref{G1vsSz}) can give a rough estimate of the  quantities in the right hand side of that equation.  

\begin{figure}[!]
\begin{center}
\includegraphics[height=0.8\linewidth,width=0.6\linewidth,angle=-90]{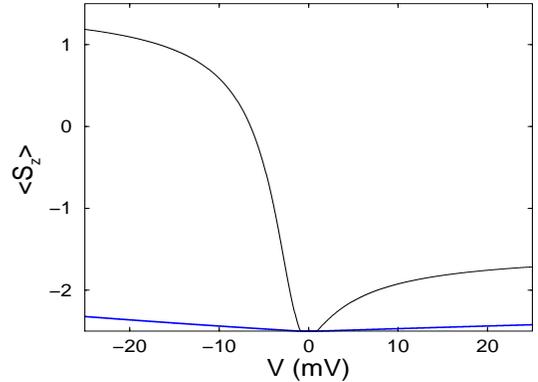}%
\end{center}
\caption{(Color online) Average magnetization $\langle S_z\rangle$ for the Mn monomer versus applied bias for $B=7$T 
corresponding to the low current (blue line) and hight current (black line) regimes.
 The tip magnetization direction was fix parallel to the easy axis and to the applied field. ${\cal P}_T=-0.31$,$T=0.5$K and $v_S=1$.}
\label{fig10}
\end{figure}
\begin{figure}[!]
\includegraphics[height=0.93\linewidth,width=0.5\linewidth,angle=-90]{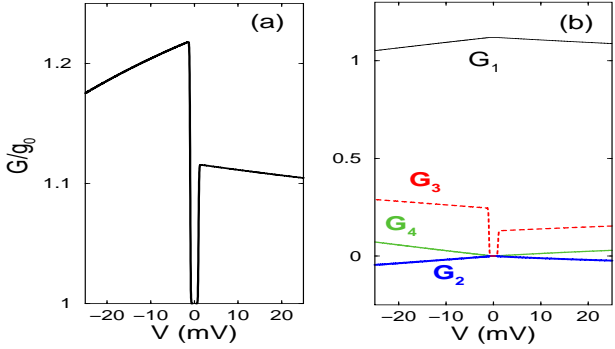}
\caption{(Color online) (a) Differential conductance for $B=7$T versus applied bias for the low current regime.
 (b) Each of the contributions to the $dI/dV$: 
$G_1$(thin-black line), $G_2$ (thick-blue line), $G_3$ (thin-dashed line)
and $G_2$ (thick-green line) versus applied bias.  $T=0.5$K, $\zeta=0.5$ and $v_T=0.7$. More detail in the text.}
\label{fig11}
\end{figure}

\subsubsection{Finite magnetic field \label{field}}
We now  analyze how  the magnetic field used in the experimental
 setup\cite{Loth_Bergmann_natphys_2010} changes the picture discussed above.
In these  experiments,  the tip polarization is achieved by attaching a single Mn atom at
the tip apex and  applying a very intense magnetic field (7T) perpendicular to the surface. 
In the experiments with spin polarized tips, \cite{Loth_Bergmann_natphys_2010} the $dI/dV$ curve changes radically from low current to high current .  In both cases, the $dI/dV$ curves are not even with respect to bias, as expected from the discussion in the previous section.  However, the lineshapes differ significantly 
 from the ones described in our previous work\cite{Delgado_Palacios_prl_2010}
 and in the previous section (Fig.~\ref{fig9}). The origin of this discrepancy can be traced back to the effect of the magnetic field on the surface adatom, absent in our previous calculation.   Since $k_bT<< g \mu_B B$, the applied field already polarizes completely the atomic spin at zero bias.
  In particular, this means that the occupation of the atomic spin state  $S_z=-5/2$ is very close to 1 and
at zero bias the average  spin of the adatom is finite, negative, and parallel to the tip spin polarization, which {\em maximizes} the $G_1$ term in the conductance.

\begin{figure}[!]
\begin{center}
\hspace{-0.9cm}
\includegraphics[height=0.93\linewidth,width=0.5\linewidth,angle=-90]{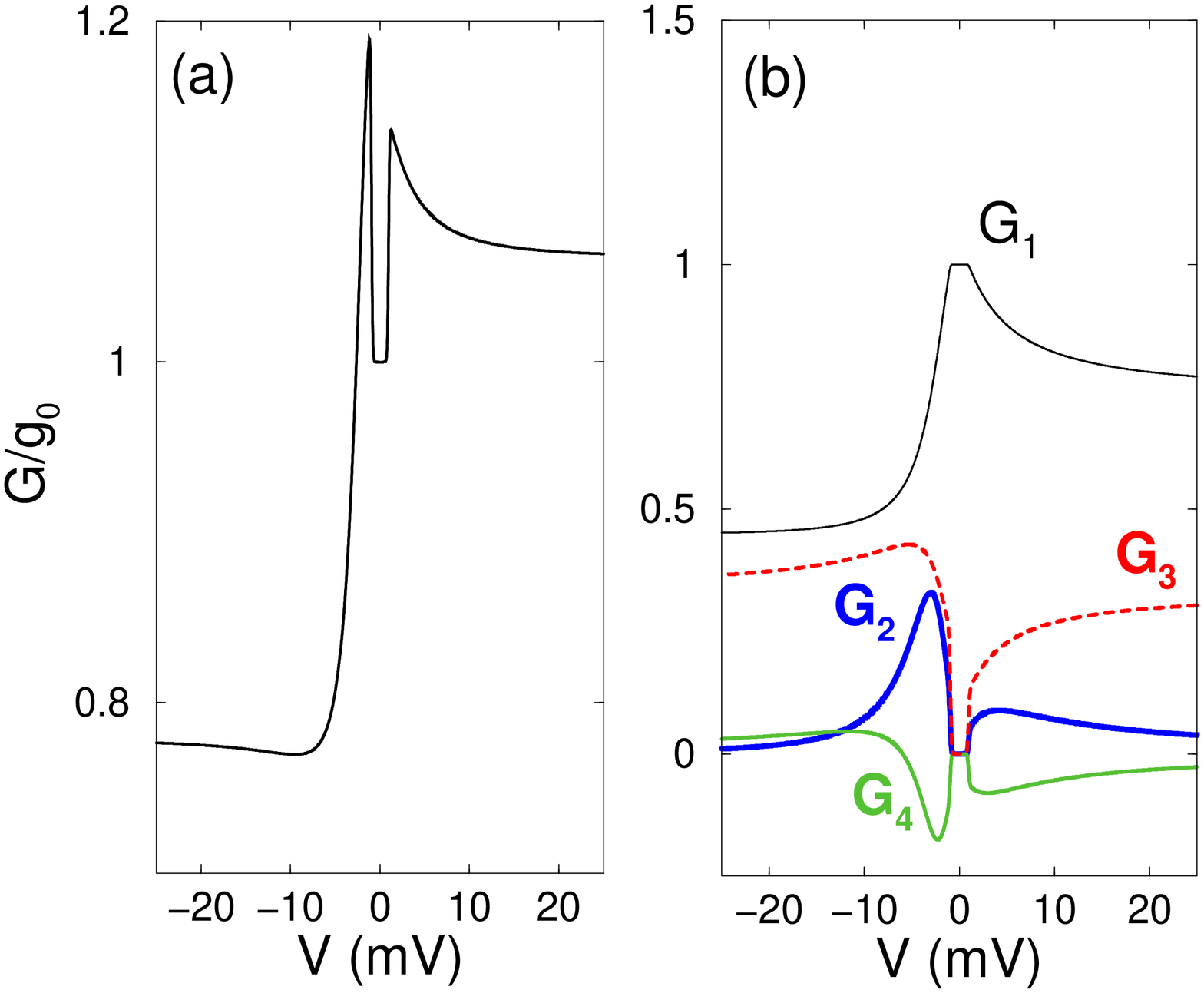} 
\end{center}
\caption{(Color online) (a) Differential conductance for $B=7$T versus applied bias for the hight current regime.
 (b) Each of the contributions to the $dI/dV$: 
$G_1$(thin-black line), $G_2$ (thick-blue line), $G_3$ (thin-dashed line)
and $G_2$ (thick-green line) versus applied bias.  $T=0.5$K, $\zeta=0.5$ and $v_T=0.7$. More detail in the text.}
\label{fig12}
\end{figure}

When  bias is applied, the occupation of the ground state $S_z=-5/2$ is depleted, which degrades the elastic contributions to the conductance. This is reflected in the evolution of the average atomic spin as a function of bias, shown in Fig.~\ref{fig10} for two cases: low current ($v_T=0.08$, $\zeta=0.44$) and high current($v_T=0.7$, $\zeta=0.5$).  The values of $v_T$ and $\zeta$, as well as the value of ${\cal P}_T=0.31$, where chosen to reproduce the experimental conductance data of Loth {\em et al.}\cite{Loth_Bergmann_natphys_2010}.   It is apparent that in both cases the depletion of the average spin is  larger for negative bias, for which transport electrons tend to counter-polarize the tip,  than for positive bias, for which the depletion can be interpreted as non-equilibrium heating of the atomic spin.   In the large current case, the current is able to reverse the average atomic spin from the equilibrium $-2.5$ at zero bias to $+1$ at -20 $mV$.   The corresponding change in the elastic components of conductance, $G_1$ and $G_2$, due to spin contrast, is shown in figures (\ref{fig11}b) for low current  and in (\ref{fig12}b) for high current.  In the high current case, these contribution are dominant, and explain the experimental  observation that the conductance is smaller at negative voltage.  Thus, in the high current case the fact that $G(+20)< G(-20)$ can be linked to the reversal of the atomic spin at negative bias and the resulting {\em reduction} of the conductance, compared to zero bias.

In the low current case the small changes in the atomic spin make $G_1$ and $G_2$, the elastic magnetoresistive contributions,  quite independent of bias, so that  change as a function bias so that they simply offset the total conductance. 
This results in a conductance which is minimal at zero bias and the  larger at large negative  bias  than at large positive bias.  
As opposed to the high current case, where the asymmetry is dominated by $G_1$ and $G_2$,  reflecting the reversal of the average atomic spin, in the low current case the asymmetry comes mostly from the asymmetry  in the  inelastic step $G_3$, associated to the term proportional to ${\cal P}_T$.

\subsection{Universal magnetization profile\label{universal}}
In Fig.~\ref{fig3pol}(c) we plot the saturation magnetization, reached after a large enough bias is applied, 
as  a function of  tip polarization. We have verified that   this curve turns out to be independent of all transport parameters, $\zeta$, $r$, $v_\eta$... and it depends only on the spin of the magnetic atom  $S$ and anisotropy parameters $D$ and $E$. 
In fact, it does not depends either on the temperature, 
as long as our approach of neglecting the phonon contribution remains valid. When combined with eq. (\ref{G1vsSz}), 
this could be used to  determine the tip polarization.

The universality of the saturation atomic magnetization comes from the fact that, at large bias, 
$i_\pm(\Delta_{M,M'}+eV)\approx  \pm eV$ so that the equation (\ref{ratenew})  for the rates can be simplified to:

\beqa
W_{M,M'} &\propto& V\left[
|S_z^{M,M'}|^2
+\frac{\rho_{T\uparrow}}{\rho_T}|S_-^{M,M'}|^2+\frac{\rho_{T\downarrow}}{\rho_T}|S_+^{M,M'}|^2
\right].
\cr 
&&
\label{wleads}
\eeqa 
Making use of Eq. (\ref{wleads}), the master equation for the occupation of the spin states in steady state 
reads 
\beqa
\sum_{M'}' &&\left[|S_z^{M,M'}|^2
+\frac{{\cal P}_T+1}{2}|S_-^{M,M'}|^2+\frac{1-{\cal P}_T}{2}|S_+^{M,M'}|^2
\right]
\cr
&&\hspace{1.cm}\times \left(P_{M'}-P_M\right)=0,\qquad \forall \;M,
\label{masteru}
\eeqa
where the prime indicates sum is done over $M'\ne M$.
Eq.~(\ref{masteru}) shows that, in the large bias limit, 
 the atomic spin steady state occupations $P_M$, and consequently the average magnetization $\langle S_z\rangle$,
   depend only on the matrix
elements of the spin operators,  and the polarization of the tip, and do not depend on the coupling strength to the tip and surface.

\section{Summary and Conclusions\label{conclusions}}

We have studied the mutual influence of non-equilibrium transport electrons and  the spin of  one and two magnetic adatoms  in  STM configuration.  Our results indicate that   non-equilibrium effects are essential to understand present IETS STM spectra of magnetic  adatoms.   
Our theory is able to describe correctly the experimental observations of IETS on single Mn atoms, both with non-magnetic \cite{Hirjibehedin_Lin_Science_2007}and magnetic\cite{Loth_Bergmann_natphys_2010}  tips,  on Mn dimers  at low  and high current\cite{Loth_Bergmann_natphys_2010},  and on FePc molecules\cite{Tsukahara_Noto_prl_2009}.  Our theory is based on a phenomenological spin model to describe both  the atomic spin states and their coupling to the transport electrons.  Current is calculated to lowest order in the tunneling Hamiltonian and depends on the occupation of the spin states $P_M$.     The spin dynamics is described with a master equation for the $P_M$  that includes the effect of carrier induced spin relaxation and the spin pumping due inelastic spin-flip assisted tunneling.   

When the time elapsed between inelastic spin flip events, $T_q$ is much longer than the atomic spin relaxation time, $T_1$,  the atomic spin remains in equilibrium and $P_M$ does not depend on bias.  In that limit, the differential conductance $dI/dV$ is piecewise constant\cite{Rossier_prl_2009} except for the steps when the bias voltage matches the energy of the spin excitations.   
When the $T_q$ is comparable or smaller than $T_1$,  the  occupations of the atomic spin states are driven away from equilibrium and they depend on the bias voltage.   Whereas non-monotonic  $dI/dV$ had been already observed experimentally, the recent results reported by Loth {\em et al.}\cite{Loth_Bergmann_natphys_2010}  have confirmed this scenario by controlling the tip-adatom distance. In addition, the use of spin-polarized tips amplifies the changes in the $dI/dV$ curves as the conductance is increased. 

The results of Loth {\em et al.}\cite{Loth_Bergmann_natphys_2010} also indicate that, in the case of magnetic tips,  the orientation of the average atomic spin can be switched at will from parallel ($V>0$) to antiparallel ($V<0$) with respect to the magnetic tip. The control of the spin of a single atom and a single magnetic molecule had been predicted by theory \cite{Rossier_prl_2009,Misiorny_Barnas_prb_2007} The control of atomic spin with non-equilibrium spin polarized carriers is similar to that obtained by optical pumping\cite{Gall_Besombes_prl_2009}.

The  main conclusions are the following:

\begin{enumerate}
\item The dynamics of the atomic spin under the influence of tunneling electrons is governed by  two intrinsic time scales, the inelastic transport time $T_q$ and the atomic spin relaxation time $T_1$. When the current induced spin flips occur more often than the time it takes to the atomic spin to relax, i.e., when $T_q<T_1$, non-equilibrium effects build-up. This makes  the occupation of the spin states different from that of equilibrium.

\item The conductance lineshape is sensitive to the occupation of the atomic spin states. This might be used to perform transport-detected single atom resonance experiments.  Our calculations indicate that non-equilibrium effects have been observed in single Mn atoms, in Mn dimers and in FePc molecules. 

\item{ A rough estimate of the quality factor of the spin excitation with energy $\Delta=\hbar\omega$, defined as
$Q=\omega T_1$,  can be obtained from the transport experiments.  We assume that the current at which the low temperature  conductance line shape starts to deviate from a piecewise constant function yields  $T_q\simeq T_1$.  Let $\Delta G_{\rm in}$ and $V$ be the height and bias of the  primary inelastic step. Then, the inelastic current is  $I_{\rm IN} = \Delta G_0 V=e/T_q$. 
Then we get
\begin{equation}
\omega T_1\simeq  \frac{G_0}{\Delta G}.
\end{equation}
}
\item Spin polarized STM can be used to magnetize the atomic spin both parallel or antiparallel to the magnetic tip moment. When electrons tunnel through the magnetic atom  from the magnetic tip to the surface,  the atomic is magnetized parallel to the tip. Reversing the bias results in an opposite spin polarization. 

\item The bias induced adatom spin polarization results in asymmetric   conductance lineshapes due, in most part, to the dependence of conductance on the relative orientation of the adatom and tip magnetizations.

\item The saturation atomic spin magnetization, obtained at large 
bias,   is only a function of the impurity spin, the anisotropy parameters and the polarization of the tip.

\end{enumerate} 

Future work should address open problems, like the origin of the spin assisted tunneling Hamiltonian for spin larger than $1/2$,  the effect of atomic spin coherence, which we have  neglected in the master equation (\ref{master}) and the fact that the observed inelastic steps in the Mn dimer do not depend on the lateral position of the tip, in contrast with our theory.

\begin{center}{\bf ACKNOWLEDGMENT }\end{center}

We acknowledge fruitful discussions with A. S. N\'u\~nez,
J. J. Palacios, C. F. Hirjibehedin and S. Loth.  This work has been financially supported by MEC-Spain (Grants JCI-2008-01885, MAT07-67845  and  CONSOLIDER CSD2007-00010).

\appendix

\section{Equation for the rates}
%
In this appendix we derive the general expression of the transition rates. Applying the Fermi Golden rule using the tunneling Hamiltonian (\ref{HTUN}) as perturbation, one gets
\beqa
\Gamma_{k\sigma M,k'\sigma 'M'}^{\eta \to \eta'}&=&\frac{2\pi}{\hbar}\delta\left(\epsilon_{\sigma\eta}(k)+E_M-\epsilon_{\sigma'\eta'}(k')-E_{M'}\right)
\crcr
&\times&
\left|\sum_{\alpha,\sigma\sigma'} T_\alpha \frac{\tau_{\sigma\sigma'}^{(\alpha)}}{2} 
{\bf S}_{\alpha,\eta\eta'}^{M,M'}   \right|^2.
\label{qrates0}
\eeqa

The modulus square in Eq. (\ref{qrates0}) can be expanded to obtain
\beqa
\sum_{\alpha,\beta,\sigma\sigma'}T_{\alpha} T_{\beta} \frac{\tau^{(\alpha)}_{\sigma\sigma'}}{2}
\frac{\tau^{(\beta)}_{\sigma'\sigma}}{2}
{\bf S}_{\alpha,\eta\eta'}^{M,M'}  {\bf S}_{\beta,\eta\eta'}^{M',M}.
\label{modulus}
\eeqa
Considering the explicit form of the Pauli matrix elements, the sum over $\sigma$ and $\sigma'$ can be done, with
just a few non-zero contributions. Using the definition of the transition rates $W_{M,M'}^{\eta\to\eta'}$, 
Eq. (\ref{frates}), we obtain after some algebra
\begin{eqnarray}
W_{M,M'}^{\eta\to\eta'}&=& 
= \frac{2\pi T_0^2\chi^2}{\hbar}
{\cal G}(\Delta_{M,M'}
+\mu_\eta-\mu_{\eta'})
\Sigma_{M,M'}^{\eta\eta'},
\label{ratenew}
\end{eqnarray}
where  
\begin{eqnarray}
\Sigma_{M,M'}^{\eta\eta'}&=&\frac{1}{4}\delta_{MM'}
\left({\cal R}^+(\eta\eta')+2\zeta{\cal R}^-(\eta\eta') \sum_a
{\bf S}_{a,\eta\eta'}^{M,M}\right)
\crcr
&&\hspace{-0.5cm}+\zeta^2\Bigg[\left|{\bf S}_{+,\eta\eta'}^{M,M'} \right|^2 \rho_{\eta\downarrow}\rho_{\eta'\uparrow} +
\left|{\bf S}_{-,\eta\eta'}^{M,M'}\right|^2\rho_{\eta\uparrow}\rho_{\eta'\downarrow}
\label{sigma}
\crcr
&&\hspace{1cm}+
{\cal R}^+(\eta\eta')\left|  {\bf S}_{z,\eta\eta'}^{M,M'}  \right|^2\Bigg].
\end{eqnarray}
Here we have introduced
${\cal R}^\pm(\eta\eta')=\rho_{\eta\uparrow}\rho_{\eta'\uparrow}
\pm\rho_{\eta\downarrow}\rho_{\eta'\downarrow}$ and the operators ${\bf S}_\pm ={\bf S}_x\pm i{\bf S}_y$.
Notice that expression (\ref{sigma}) is valid for all states $M,\; M'$, in contrast to Eq. (5) of
Ref. \onlinecite{Delgado_Palacios_prl_2010} where only the inelastic matrix elements were explicitly written.
%
%

\section{Equations for the current}
Here we shall derive expressions (\ref{ielast}) and (\ref{iinelast}).
Let us start with the expression for the current, Eq. (\ref{current0}). Dividing the contribution into its elastic and inelastic
part, with the help of Eq. (\ref{ratenew}) and (\ref{sigma}), we can write
\beqa
\label{A1}
I_0+I_{MR}&=&\frac{2\pi eT_0^2}{\hbar} \frac{\chi^2\rho_S\rho_T}{8}
 \left[1+ 2\zeta  \langle{\bf S}_{z,T,S}\rangle {\cal P}_T\right]
\crcr 
&&\times
\left({\cal G}(eV)+{\cal G}(-eV)\right),
\eeqa
or, using the definition of $g_0$ and $i_+$
\beqa
I_0+I_{MR}&=&\frac{g_0}{e}
 \left[1+ 2\zeta \langle  {\bf S}_{z,T,S}\rangle {\cal P}_T\right]i_-(eV),
\nonumber
\eeqa
which is the result of Eq. (\ref{ielast}). 
For the inelastic contribution ($M\ne M'$), the difference $W_{MM'}^{S\to T}-W_{M,M'}^{T \to S}$ can be written with the help of
Eqs. (\ref{ratenew}) and (\ref{sigma}) as
\beqa
&&W_{MM'}^{T\to S}-W_{M,M'}^{S \to T}=\zeta^2\frac{2\pi\chi^2\rho_S\rho_T T_0^2}{8\hbar}\Big( i_-(\Delta_{M,M'}+eV)
\crcr
&&\hspace{0.5cm}\times
\sum_{a}\left| {\bf S}_{a,T,S}^{M,M'}\right|^2+ {\cal P}_T i_+(\Delta_{M,M'}+eV)
\crcr
&&\hspace{0.8cm}\times2{\rm Im}\left[{\bf S}_{x,T,S}^{M,M'}{\bf S}_{y,T,S}^{M',M} \right]\Big).
\nonumber
\eeqa
Using the definitions of $g_s$ and ${\bf \Xi}_{xy}(M,M')$, expression (\ref{iinelast}) is recovered. 



\bibliographystyle{apsrev}

\end{document}